\documentclass[submission, Phys]{SciPost}
\usepackage{amsmath,amsfonts,amssymb,graphicx,mathrsfs}
\usepackage{cite}
\usepackage{overpic}
\usepackage{amsbsy}
\usepackage{epsfig}
\usepackage{wrapfig}
\usepackage{array}
\usepackage[utf8]{inputenc}
\usepackage{textcomp}
\usepackage{color}
\usepackage{braket}
\usepackage{bm,hyperref}
\usepackage[titletoc,toc,title]{appendix}
\usepackage[normalem]{ulem}
\usepackage{float}
\usepackage{todonotes}

\definecolor{myblue}{rgb}{.93, .93, 1}

\newcommand{\beq}{\begin{equation}}
\newcommand{\eeq}{\end{equation}}

\newcommand{\be}{\begin{equation}}
\newcommand{\ee}{\end{equation}}
\newcommand{\mac}{\mathcal}

\newcommand{\ba}{\begin{eqnarray} }
\newcommand{\ea}{\end{eqnarray} }

\begin{document}

\begin{center}{\Large \textbf{
Fractonic critical point proximate to a higher-order topological insulator: How does UV blend with IR?}}\end{center}

\begin{center}
Yizhi You\textsuperscript{1},
Julian Bibo\textsuperscript{2},
Taylor L. Hughes\textsuperscript{3},
Frank Pollmann\textsuperscript{2}
\end{center}

\begin{center}
    {\bf 1} Department of Physics, Princeton University, NJ, 08544, USA
\\
{\bf 2} Department of Physics, Technical University of Munich, 85748 Garching, Germany
\\
{\bf 3} Department of Physics and Institute for Condensed Matter Theory, University of Illinois at Urbana-Champaign, Illinois 61801, USA
\\
\end{center}

\begin{center}
\today
\end{center}

\section*{Abstract}
{\bf
We propose an unconventional topological quantum phase transition connecting a higher-order topological insulator (HOTI) and a featureless Mott insulator sharing the same symmetry patterns. We construct an effective theory description of the quantum critical point (QCP) by combining a bosonization approach and the coupled-stripe construction of 1D critical spin ladders. 
The phase transition theory is characterized by a critical dipole liquid theory with subsystem $U(1)$ symmetry whose low energy modes contain a Bose surface along the $k_x,k_y$ axis. Such a quantum critical point manifests fracton dynamics and the breakdown of the area law entanglement entropy due to the existence of a Bose surface.
We numerically confirm our findings by measuring the entanglement entropy, topological rank-2 Berry phase, and the static structure factor throughout the topological transition and compare it with our previous approach obtained from the percolation picture. 
A significant new element of our phase transition theory is that the infrared~(IR) effective theory is controlled by short wave-length fluctuations with peculiar UV-IR mixing.}

\section{Introduction and background}

Throughout the last few decades, our understanding of quantum criticality and phase transitions has significantly advanced, especially with respect to critical points that cannot be satisfactorily characterized by spontaneous symmetry breaking within the Landau-Ginzburg-Wilson~(LGW) paradigm. In particular, continuous quantum phase transitions can appear between two distinct phases with the same symmetry pattern, but different topological properties. In this work, we propose a new type of quantum critical point~(QCP) that connects a 2D HOTI phase~\cite{you2019multipolar,bibo2019fractional,you2020fracton} and a trivial Mott insulator phase which eludes the LGW paradigm.

A new class of symmetry protected topological insulators, dubbed Higher-Order Topological Insulators~(HOTI) has been discovered in the recent literature~\cite{benalcazar2017quantized,schindler2017higher,langbehn2017reflection,song2017d}. HOTIs admit gapped surfaces separated by gapless corners/hinges where the surfaces intersect, and exemplify a rich bulk-boundary correspondence. Aside from HOTIs generated by topological band structures, recent research suggests strongly interacting bosonic systems can potentially host a HOTI having robust bosonic corner zero-modes~\cite{you2018higherorder,dubinkin2018higher,rasmussen2018classification}. A variety of mathematical invariants, topological responses, and field theory approaches~\cite{rasmussen2018classification,you2019multipolar,you2019higher} have been explored to characterize and classify HOTI band theories and interacting HOTIs, while the phase transitions between distinct HOTI states are still under exploration. In Ref.~\cite{you2020fracton}, we demonstrated the existence of a topological phase transition between an interacting HOTI phase and a trivial Mott insulator phase, whose universality does not fit into the LGW paradigm. A particularly interesting question in this regard is whether the critical region inherits the topological properties of the proximate HOTI, and how such a phase transition is influenced by the topological structure or entanglement pattern of the adjacent HOTI phase. 

However, a detailed understanding of such HOTI phase transitions starting from concrete microscopic models is often hindered by the overarching problem of the limitations of analytical and numerical approaches to deal with strongly interacting many-body systems. In our previous attempt\cite{you2020fracton}, we proposed that the QCP can be understood as the bulk percolation~\cite{chen2013critical} of domain walls that act as phase boundaries between regions of the HOTI phase and regions of a trivial Mott insulator. The corners and rough patches of the 1D domain walls can then be treated as the corners of the HOTI phase, and are thus each decorated with a robust spinon zero mode. At the QCP, the proliferation of domain walls triggers the fluctuations of the corner spinon zero modes, and precipitates fracton dynamics of the spinons that are constrained by subsystem $U(1)$ symmetry. Hence, the critical point contains quasiparticles with fracton behavior and sub-dimensional kinetics where the spinon-dipoles move only transverse to their dipole moment~\cite{Vijay2015-jj,Vijay2016-dr,Chamon2005-fc,Haah2011-ny,yoshida2013exotic,pretko2020fracton}. Based on these observations, we concluded that the phase transition is characterized by a critical dipole liquid~\cite{you2019emergent,xu2007bond,paramekanti2002ring,tay2010possible,mishmash2011bose,dubinkin2020higher,seiberg2020exotic} exhibiting exotic features including: (i) a Bose surface~\cite{sachdev2002scratching} having zero energy states that form nodal-lines along the $k_x$ and $k_y$ axes,  (ii) a breakdown in the area law of the entanglement entropy at the QCP, which is replaced by a scaling with a logarithmic enhancement $L \ln(L)$\cite{lai2013violation} ($L$ being the system size) at the critical point instead, and (iii)  `UV-IR mixing'~\cite{seiberg2020exotic,karch2020reduced,xu2007bond,paramekanti2002ring} such that the low energy part of the spectrum contains subextensive number of high momentum modes so the critical theory is dominated by short-wave length physics.
In addition to the analytic approaches, in Ref.~\cite{you2020fracton} we also used the density matrix renormalization group (DMRG) method to analyze the dipole stiffness\cite{dubinkin2019theory} together with structure factor at the QCP, both of which provided strong corroborating evidence of a critical dipole liquid with a Bose surface.

In this paper, we provide a complementary approach to tackle the HOTI transition from an alternative perspective which gives a more controlled way to access the microscopic physics compared to the intuitive percolating domain wall picture from Ref.~\cite{you2020fracton}. To achieve this we will begin with a quasi-1D limit of our microscopic model by placing the Hamiltonian on a one-dimensional spin ladder (stripe). As the HOTI phase we propose only requires subsystem U(1) and $\mathcal{T}$ symmetries, such a thin-stripe limit, which breaks the $C_4$ crystalline symmetry, does not interfere with the robustness of the corresponding HOTI state. Combining this with bosonization, we show that the quantum critical point between the HOTI and trivial Mott phase defined on the 1-dimensional stripe is described by a $c=1$ conformal field theory (CFT) with critical dipoles fluctuating along the stripe. We verify the validity of this theory by measuring the entanglement entropy and the rank-2 topological Berry phase\cite{you2019multipolar,dubinkin2019theory} using DMRG and exact diagonalization techniques. Further, we implement a coupled stripe construction to recover the original microscopic model Hamiltonian proposed in Ref.~\cite{you2020fracton} and restore the 2D critical theory. We hence confirm that the critical point can be interpreted as a 2D critical dipole liquid theory with a Bose surface and algebraically decaying dipole correlations\cite{you2019emergent,xu2007bond,paramekanti2002ring,tay2010possible,mishmash2011bose,dubinkin2020higher,seiberg2020exotic}. Perhaps surprisingly, the inter-stripe coupling between 1D critical ladders does not open a gap, nor trigger any superfluid instability, when approaching the 2D limit due to the subsystem symmetry protection. Instead, it merely engenders dipole hopping and coherence between the transverse stripes which results in a Bose surface with zero energy modes along the $k_x$ and $k_y$ axes. 
This conclusion exactly matches our previous work\cite{you2020fracton}, where we adapted a non-perturbative approach with a percolating domain wall picture to describe the theory for the critical point.

\section{Higher order topological Mott insulator}\label{section1}

In this section, we review a class of 2D gapped bosonic models that host gapless corner modes\cite{you2020fracton,you2018higher,you2018higherorder}, which  are protected by the combination of $U(1)$ subsystem symmetry, and time-reversal $\mathcal{T}$. This phase can be interpreted as a higher order topological Mott insulator with quantized topological quadrupole moment $Q_{xy}$\cite{you2018subsystem,you2018symmetric,devakul2018strong,devakul2018fractal,you2019multipolar}. 

\begin{figure}[h]
\centering
\includegraphics[width=0.8\textwidth]{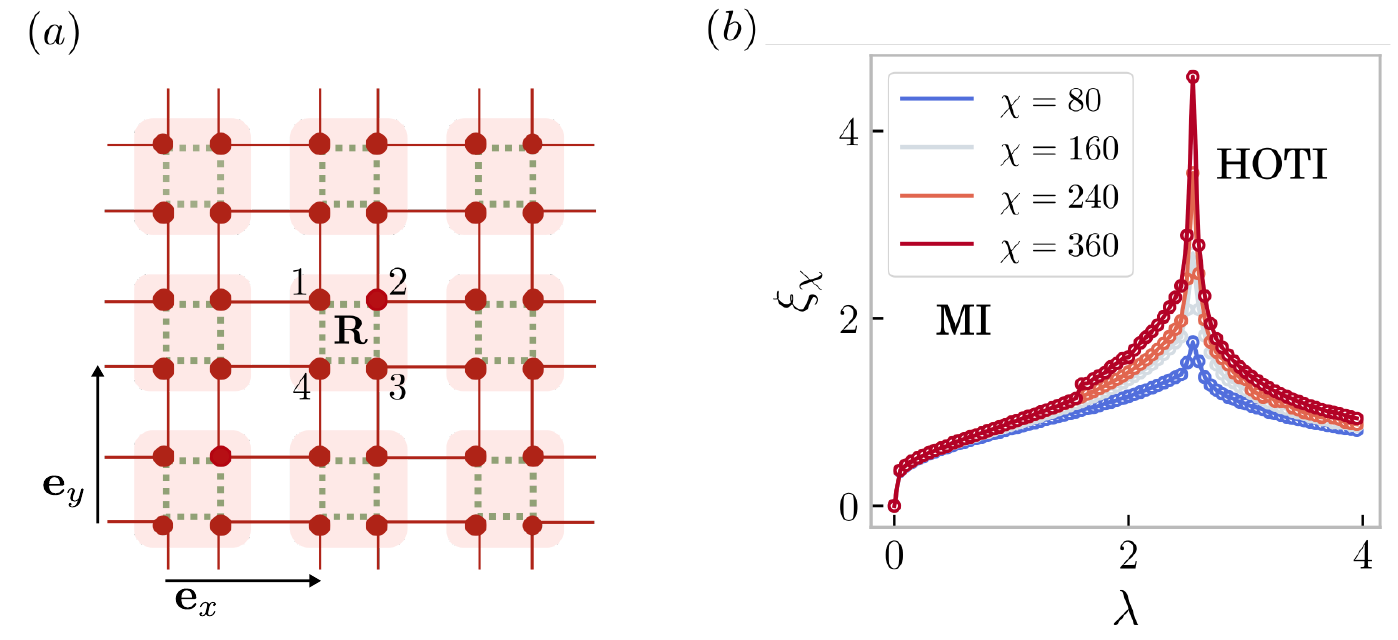}
\caption{\textbf{Lattice model and quantum critical point (QCP) from from Ref.~\cite{you2020fracton}.} Panel $(a)$ shows the lattice model. Each unit cell consists of four spin $S=1/2$ (red dots). The ring-exchange terms (red squares) couple four neighboring unit cells, while sites in the unit cell are isotropically coupled via an $XY$ interaction (green dotted lines). Panel (b) shows the correlation length $\xi_\chi$ as function of the tuning parameter $\lambda$ and the bond dimension $\chi$ for an infinite cylinder along $x$, $L_x=\infty$, and periodic boundary conditions along $y$ with a circumference $L_y =6$. A second order quantum phase transition between a Mott-Insulator (MI) and higher order topological insulator (HOTI) occurs at $\lambda_c \approx 2.54$. } 
\label{phase}
\end{figure}

To frame our discussion, we consider the following model on a 2D square lattice with four spin-1/2 degrees of freedom per unit cell as shown in Fig.~\ref{phase}(a):  
\begin{align}  
&H=H_{XY}-\lambda H_{\text{ring exchange}}\nonumber\\
&= \sum_{{\bf{R}}}\left(S^{+}_{{\bf{R}},1} S^{-}_{{\bf{R}},2}+S^{+}_{{\bf{R}},2} S^{-}_{{\bf{R}},3}\right.
\left.+S^{+}_{{\bf{R}},3} S^{-}_{{\bf{R}},4}+S^{+}_{{\bf{R}},4} S^{-}_{{\bf{R}},1}\right)\nonumber\\
&-\lambda\sum_{{\bf{R}}}\left( S^{+}_{{\bf{R}},2} S^{-}_{{\bf{R}}+\mathbf{e}_x,1} S^{+}_{{\bf{R}}+\mathbf{e}_x+\mathbf{e}_y,4} S^{-}_{{\bf{R}}+\mathbf{e}_y,3}\right)+h.c.
\label{hhh}
\end{align}
This Hamiltonian contains an inter-cell ring-exchange interaction between the four spins located at the four corners of each red plaquette, and an XY spin interaction within the unit cell as shown in Fig.~\ref{phase}(a). In passing we mention that models having ring-exchange terms of this type can be realized in cold-atom settings, which hence is a natural arena for experimental investigation of our subsequent predictions~\cite{paredes2008,dai2017}. The magnon creation/annihilation operators $S^{\pm}=\sigma^x\pm i\sigma^y$ can be mapped to a hardcore boson description using $b^{\dagger}(b)=S^+(S^-), S^z=n_b-1/2$, where $n_b=b^{\dagger}b^{\phantom{\dagger}}$; we use both languages interchangeably where convenient.

Eq.~\eqref{hhh} exhibits time-reversal symmetry $\mathcal{T}=\prod_{\bf{R}}\prod^{4}_{m=1}i\sigma^y_{\mathbf{R},m}\mathcal{K}$  and a subsystem U(1) symmetry that conserves the sum of $S^z$ inside the unit cell $\mathbf{R}$, i.e. $S^z(\mathbf{R})=\sum^4_{m=1}S^z_{\mathbf{R},m},$ for every row (R) and column (C),
\begin{equation} 
U^{\mathrm{sub}}_{R(C)}(1): \prod_{\mathbf{R} \in R(C)} e^ {i \theta  S^z(\mathbf{R})}.
\end{equation} 
Under subsystem symmetry, any spin-bilinear XY coupling is forbidden \emph{between} unit cells. Furthermore, the global $\mathcal{T}$ symmetry forbids terms in the Hamiltonian that  polarize the spins, and would be spontaneously broken in the presence of magnetic order.
The subsystem U(1) symmetry restricts the mobility of the magnon, hence the leading order dynamics is attributed to pairs of dipoles, each composed of a particle-hole pair on a lattice link, that hops along the direction transverse to its dipole moment. 

To gain intuition for this model, we can interpret the model in terms of hardcore bosons  as:
\begin{align}  
H=& \sum_{{\bf{R}}}\left(b^{\dagger}_{{\bf{R}},1} b_{{\bf{R}},2}+b^{\dagger}_{{\bf{R}},2} b_{{\bf{R}},3}\right.
\left.+b^{\dagger}_{{\bf{R}},3} b_{{\bf{R}},4}+b^{\dagger}_{{\bf{R}},4} b_{{\bf{R}},1}\right)\nonumber\\
\hphantom{=}&-\lambda\sum_{{\bf{R}}}\left( b^{\dagger}_{{\bf{R}},2} b_{{\bf{R}}+\mathbf{e}_x,1} b^{\dagger}_{{\bf{R}}+\mathbf{e}_x+\mathbf{e}_y,4} b_{{\bf{R}}+\mathbf{e}_y,3}\right)+\text{h.c}.
\label{hhhboson}
\end{align}
In this language, the subsystem U(1) symmetry operator implements a (possibly different) phase rotation for the bosons in each row/column,
\begin{equation} 
U^{sub}(1): b_j \rightarrow e^{i\theta } b_j,\;\; j \in \text{row}.
\end{equation} Additionally, $\mathcal{T}$ acts as a particle-hole symmetry for the hardcore bosons,
\begin{align} 
\mathcal{T}:&|1 \rangle \rightarrow | 0 \rangle, |0 \rangle \rightarrow  - |1 \rangle \\
& b \rightarrow -b^{\dagger}, b^{\dagger} \rightarrow -b. 
\end{align}
 This Hamiltonian has a boson ring-exchange interaction among four of the hardcore bosons around each plaquette, in addition to the intra-site hopping terms. 
 
 As $\lambda$ is tuned, the system effectively has competing orders dominated by either the intra-cell XY interaction, or the inter-cell plaquette ring exchange interaction. In the limit $\lambda \sim 0$, the intra-cell term plays the key role and generates an entangled cluster within each unit cell. The resulting ground state, which is simply a tensor product of the ground state of each unit cell, is a featureless Mott insulator with a magnon gap for both the bulk and boundary.
In the opposite limit $\lambda \sim \infty$, it was shown in Ref.~\cite{you2019multipolar} that
the plaquette term projects the four interacting spins into a unique maximally entangled state
$|\downarrow_2 \uparrow_1 \downarrow_4 \uparrow_3 \rangle+|\uparrow_2 \downarrow_1 \uparrow_4 \downarrow_3\rangle$
where we omitted unit cell labels.
The corresponding ground state for the entire system is thus a product of entangled plaquettes, and has a finite magnon gap in the bulk. In the presence of a smooth boundary, each edge unit cell naively has two free spin-1/2 modes, but these can be coupled and gapped to form a 
singlet state via the onsite XY coupling in Eq.~\eqref{hhhboson} in the presence of infinitesimal $\lambda$. For rough edges and/or corners, there is an additional spin-1/2 zero mode per corner whose two-fold degeneracy is protected by $\mathcal{T}$ and subsystem U(1) symmetry. Based on this observation, this gapped ground state when $\lambda$ dominates is a higher-order topological insulator with robust corner modes protected by $\mathcal{T}$ and subsystem U(1) symmetry. 

This subsystem symmetric HOTI phase was studied in Ref.~\cite{you2019multipolar}, and was shown to exhibit a quantized quadrupole moment density of $Q_{xy}=1/2$. Since the hardcore boson model in Eq.~\eqref{hhhboson} has subsystem U(1) symmetry, we can introduce a higher rank\cite{you2019higher} (rank 2) background gauge field $A_{xy}$ that minimally couples to the ring exchange term\cite{you2018symmetric,pretko2017fracton,pretko2017subdimensional,Vijay2016-dr}:
\begin{align}  \label{Eq:H2Gauge}
&- \sum_{{\bf{R}}}\left[ b^{\dagger}_{{\bf{R}},1} b^{\phantom{\dagger}}_{{\bf{R}}+e_x,4}b^{\dagger}_{{\bf{R}}+e_x+e_y,3}b^{\phantom{\dagger}}_{{\bf{R}}+e_y,2}e^{i A_{xy}({\bf{R}})}+h.c \right].\nonumber\\
\end{align}
We denote $A_{xy}$ as the gauge field living in the center of each plaquette that couples with the dipole current $J_{xy}({\bf{R}})=(i b^{\dagger}_{{\bf{R}},1} b^{\phantom{\dagger}}_{{\bf{R}}+e_x,4} b^{\dagger}_{{\bf{R}}+e_x+e_y,3}b^{\phantom{\dagger}}_{{\bf{R}}+e_y,2}+h.c)$. The dipole current is proportional to the ring-exchange term on the plaquette, and can be viewed as a dipole oriented along $x$ hopping along $y,$ or vice versa. We also introduce a time component $A_{0}$ of the gauge field, that couples with the total charge density $n_b$. These gauge fields transform under gauge transformations $\alpha$ as
\begin{align}
    &A_{xy}\rightarrow A_{xy}+\partial_x\partial_y \alpha\nonumber\\
    &A_{0}\rightarrow A_{0}+\partial_t\alpha,
    \end{align}
from which we find a single gauge-invariant combination of these fields: 
\begin{equation}
    E_{xy} =   \partial_x \partial_y A_0 -\partial_t A_{xy}.
\end{equation}

Following the spirit of Goldstone-Wilczek (polarization)\cite{goldstone1981fractional,qi2011topological} response that is found in 1D SPTs with protected end modes, our HOTI model admits a generalization of this topological response theory, described by the quadrupolarization $\Theta$-term:
\begin{align} 
&\mathcal{L}_{Q}=\frac{\Theta}{2\pi} (\partial_x \partial_y A_0-\partial_t A_{xy})=\frac{\Theta}{2\pi}E_{xy}.
\label{qua}
\end{align}
As we will show at the end of this subsection, $\Theta$ is defined modulo $2 \pi$.  Hence even though the electric field $E_{xy} =\partial_x \partial_y A_0-\partial_t A_{xy} $ is odd under 
 $\mathcal{T}$ symmetry 
 \begin{equation*}
     \mathcal{T}: A_0 \rightarrow -A_0, A_{xy} \rightarrow  -A_{xy}, (x,y,t)\rightarrow (x,y,t)
 \end{equation*} (which here acts as a PH symmetry), 
 such a term $\mathcal{L}_Q$ is symmetry-allowed provided that $\Theta = 0, \pi$.  Hence, the action (\ref{qua}) describes the response of a subsystem protected topological order\cite{you2018subsystem,you2018symmetric}, with a coefficient $\Theta$ that is quantized in the presence of $\mac{T}$ symmetry.

We now consider the nature of the charge response described by (\ref{qua}).  The action describes the following charge and dipole current response:
\begin{align}
    j_0&=\frac{1}{2\pi}\partial_x \partial_y \Theta,~j_{xy}=\frac{1}{2\pi}\partial_t \Theta.
    \end{align} 
By comparing these equations with the expected properties of a quadrupole moment $Q_{xy}$ \cite{raab2005,benalcazar2017quantized}, we can identify the bulk quadrupole moment to be $Q_{xy}=\frac{\Theta}{2\pi}.$  Similar to the dipole moment of the 1D topological insulator, we find that in the rank-2 case, $Q_{xy}$ is quantized in the presence of symmetry (time-reversal, which acts as charge-conjugation symmetry in this context).  
 Interestingly, the action (\ref{qua}) predicts fractional charge on corners at the intersection between edges with normal vectors $\hat{x}$ and $\hat{y}.$ At such a corner, which we can heuristically model as a product of step functions in the $\Theta$ field, we have $j_0=\partial_x \partial_y \frac{\Theta}{2\pi} =\pm \tfrac{1}{2} \delta(x-x_0)\delta(y-y_0)$,  where $(x_0, y_0)$ is the position of the corner.  For $\Theta = \pi$, this indicates the presence of a half-charge localized to the corner.  
In the spin language, the corner response with half the charge of local bulk excitations corresponds to an unpaired spin-$1/2$ zero mode. 
The $\mathcal{T}$ symmetry, having effectively $\mathcal{T}^2=-1$ for the corner spin, ensures the Kramers' spin degeneracy at the corner cannot be lifted by a Zeeman field.  These results all match the expected phenomena from the microscopic model.

In our previous work~\cite{you2020fracton}, we obtained the phase diagram of the Hamiltonian in Eq.~\eqref{hhh}  using the DMRG method~\cite{White1992, Mcculloch2008, Hauschild2018} on an infinitely long cylinder as shown in Fig.~\ref{phase}(b). We find that the aforementioned exactly solvable limits extend into two gapped phases (a trivial Mott insulator and a HOTI respectively). Importantly, our numerics indicate that the two gapped phases are connected by a second-order phase transition at $\lambda_c\approx2.54$. It is noteworthy to emphasize that the HOTI phase and the trivial Mott phase display the same symmetries, but harbor distinct topological features, and thus cannot be differentiated via any local observable. This further implies that the QCP between the phases cannot be accessed by a LGW type phase transition theory based on a fluctuating order parameter. 

In our earlier approach\cite{you2020fracton}, we used a percolation picture to obtain the low energy effective theory of the quantum critical point. In this manuscript, we will try an alternative route by considering a quasi-1D stripe limit of this model. As the spatial rotation symmetry is not required for this class of HOTI, the underlying physics including the topological quadrupole response and protected corner mode will not change in the thin stripe limit.
We will demonstrate that the quantum phase transition of the Hamiltonian in Eq.~\eqref{hhh} in the thin stripe limit can be accessed by a critical dipole model on a 1D ladder with central charge $c=1$. The original quantum phase transition in the 2D limit can then be attained by a coupled ladder construction, on which we will elaborate later.





\subsection{The model and its thin stripe limit}

\begin{figure}[h]
  \centering
    \includegraphics[width=0.7\textwidth]{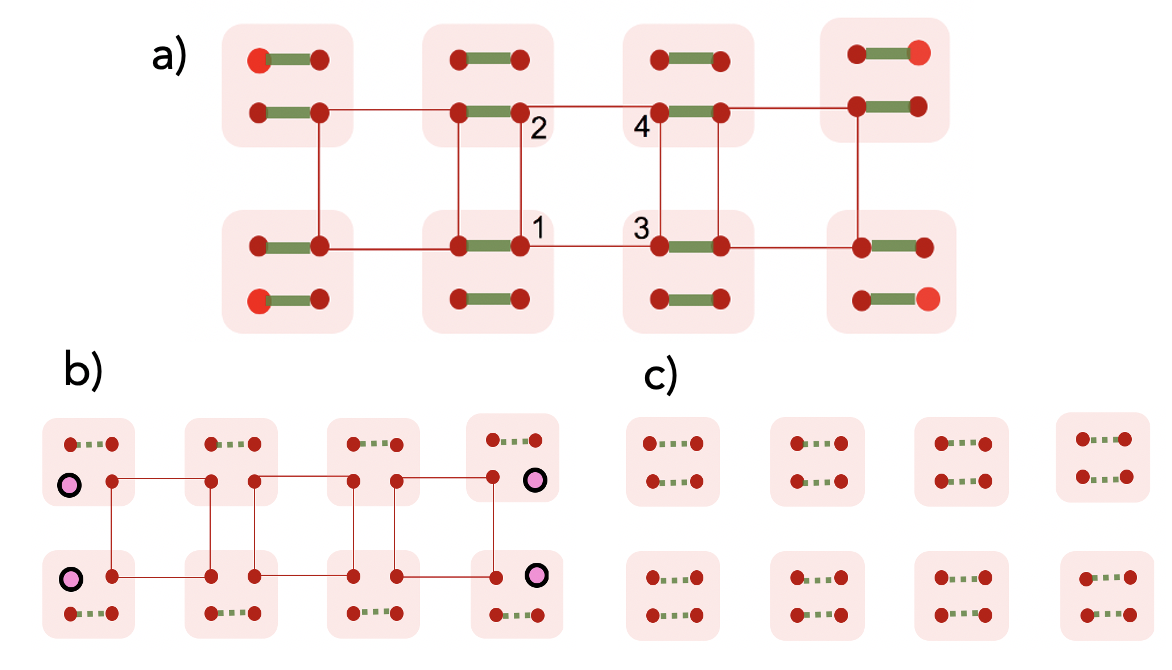} 
  \caption{a) 1D stripe model with ring exchange term (red plaquette) and intra-site XY coupling (green bond). b) The HOTI state with plaquette entangled clusters and `corner zero modes' at the boundary. c) The trivial Mott state with onsite entangled pairs.} 
    \label{case1}
\end{figure}

In this section, we begin with a 1D stripe model that can be viewed as an anisotropic limit of Eq.~\eqref{hhhboson}:
\begin{align}  
&H= \sum_{{\bf{R}}}~\left[(b^{\dagger}_{{\bf{R}},1} b_{{\bf{R}},2}+b^{\dagger}_{{\bf{R}},3} b_{{\bf{R}},4})
-\lambda'  b^{\dagger}_{{\bf{R}},2} b_{{\bf{R}}+\mathbf{e}_x,1} b^{\dagger}_{{\bf{R}}+\mathbf{e}_x+\mathbf{e}_y,4} b_{{\bf{R}}+\mathbf{e}_y,3}+h.c.\right].
\label{hhhstripe}
\end{align}
The model is defined on a spin ladder as shown in Fig.~\ref{case1} that can be treated as a single stripe of the 2D lattice from Fig.~\ref{phase}-(a). Each unit cell contains four spins with intra-site XY coupling on the x-links inside each unit cell. While the original model in Eq.~\eqref{hhhboson} also contains intra-site XY coupling on the y-links inside each unit cell, we will drop this term for now when we focus on the 1D stripe limit. We will recover this coupling when analyzing the coupled stripe construction in 2D in the forthcoming section. Our Hamiltonian also contains ring-exchange interactions among the four spins on each plaquette of the ladder. 
This model respects the $\mathcal{T}$ and subsystem U(1) symmetries we defined in our previous discussion. In particular, the charge is conserved on each row of the ladder as well as each rung between the ladder\footnote{For the thin stripe limit, the subsystem U(1) charge conservation on the y-column becomes a local charge conservation on each rung.} so any XY coupling between unit cells is prohibited.

In the limit $\lambda' \rightarrow 0$, the intra-site XY coupling dominates and projects the four spins in each unit cell onto two dimer singlets. The resultant ground state is a featureless gapped Mott insulator. When $\lambda' \gg 1$, the ring-exchange term dominates and projects the four spins on the coupled plaquette into a maximally entangled cluster state. Additionally, the remaining two dangling spins in each unit cell are paired into an onsite dimer singlet provided there is weak intra-site XY coupling. Subsequently, the ground state of this thin stripe model is fully gapped with a finite correlation length.
In the presence of an open stripe, each `corner' of the stripe contains an additional boson zero-mode that is not paired with any partner. Although the two corners on the same rung are now adjacent to each other, one cannot gap them out without breaking subsystem U(1) symmetry since the charge on the first and second rows of the stripe are conserved independently. Thus, the corner zero modes are fully protected by $\mathcal{T}$ and subsystem U(1), and the ground state renders a higher-order topological Mott insulator even in the thin stripe limit.

\begin{figure}[h]
  \centering
    \includegraphics[width=1\textwidth]{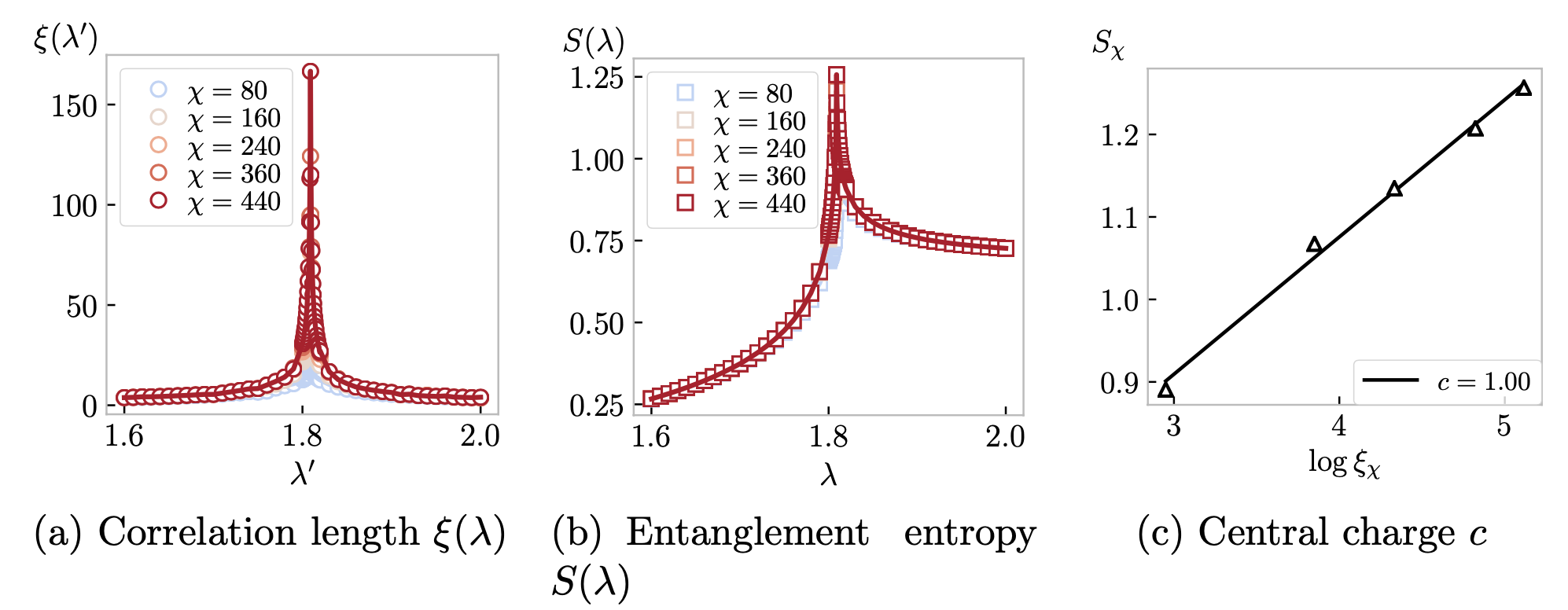} 
\caption{\textbf{Quantum critical point $\lambda_c\approx 1.809$.} The panels $(a)$ and $(b)$ show the correlation length $\xi(\lambda)$ and the half-infinite stripe entanglement entropy $S(\lambda)$ for the QCP of the model defined in Eq.~\eqref{hhhstripe} for various bond dimensions $\chi$. In the last panel the central charge of $c$ of the corresponding CFT is shown, which perfectly matches the anlytical predicted value.}
\label{fig:1dphase}
\end{figure}

\begin{figure}[h]
  \centering
    \includegraphics[width=0.5\textwidth]{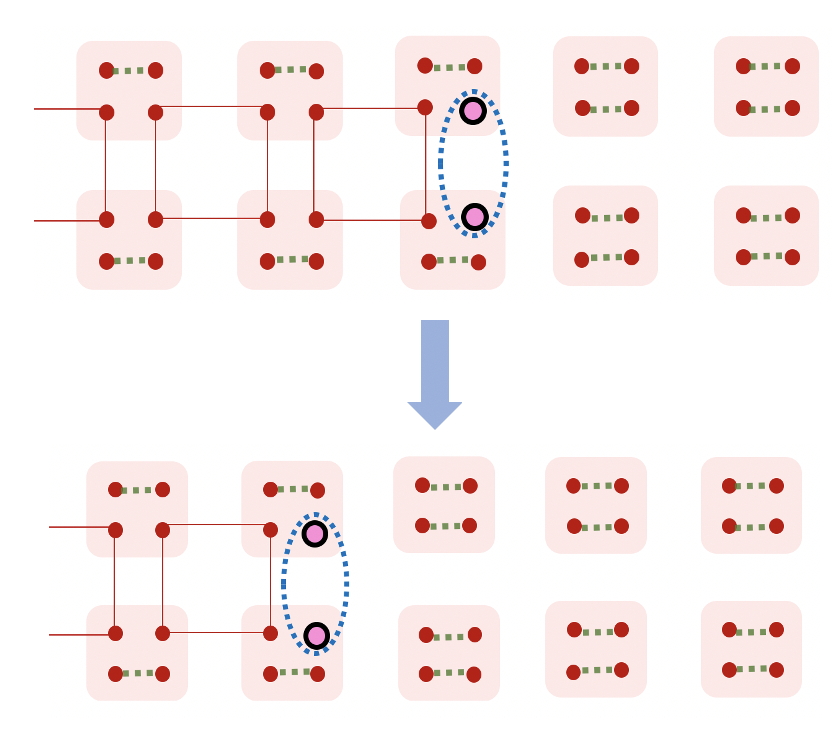} 
  \caption{Top: The domain wall between HOTI ground state (with plaquette entangled patterns) and trivial Mott state (with intra-site entangled patterns) carries a pair of boson zero modes (denote dipole) between the rung. Bottom: The spatial fluctuation of the domain wall triggers the hopping of the dipole.} 
    \label{case2}
\end{figure}

The HOTI phase exhibits patterns with plaquette entangled configurations between unit cells while the trivial Mott phase, despite sharing the same symmetry, contains dimer patterns within each unit-cell. As $\lambda'$ is tuned toward the quantum critical region, the competition between two phases generates `domain wall defects' separating the two distinct patterns with either plaquette entangled configurations or intra-site entangled dimers. Each domain wall, as illustrated in Fig.~\ref{case2}, contains a pair of boson zero modes (denoted as a dipole) between the rungs. These two bosons from different rows cannot be paired due to the subsystem U(1) symmetry and $\mathcal{T}$ symmetry constraints. At the critical point, the fluctuation of the domain walls enables the spatial fluctuation of dipole hopping along the stripe. In the meantime, a single boson from each row cannot fluctuate alone so the leading order dynamics is still controlled by boson-pair (dipole) hopping term. 

We now provide both analytic argument and numerical evidence to demonstrate that the quantum critical point separating the HOTI phase and the trivial Mott phase on the thin stripe is akin to the 1D critical boson theory with central charge $c=1$. Our DMRG simulations indicate the quasi-1D Hamiltonian in Eq.~\eqref{hhhstripe} has a continuous phase transition between the HOTI phase and trivial Mott phase when tuning $\lambda'$. In Fig.~\ref{fig:1dphase}(a), we identify the quantum critical point $\lambda'_c\approx 1.809$ between two phases with a divergent correlation length. Hence we find that the transition occurs when $\lambda'$ is comparable to the intra-cell tunneling strength. In this regime the plaquette entangled patterns and on-site entangled patterns compete and coexist on the stripe. The coexistence and spatial phase separation can be viewed from a percolation picture illustrated in Fig.~\ref{case2}, and discussed in detail in Ref. \cite{you2020fracton}. In the quantum critical region adjacent to the trivial Mott phase, the plaquette ring-exchange term triggers some regions containing plaquette entangled states that exhibit the HOTI ground state pattern. Domain walls that form between the two phases can be viewed as the `boundary' between the HOTI and trivial phases on the stripe, and hence they harbor a pair of boson zero modes, i.e., a dipole between the rungs. In the quantum critical region, strong fluctuations between different phase separation patterns are induced, and the domain walls tend to proliferate. These spatial fluctuations concurrently trigger the dynamics of the dipole hopping along the ladder with the domain wall, similar to the percolation of domain wall defects in 1D critical systems~\cite{chen2013critical}. 

As the boson zero mode inside the domain wall is a spinon, we can represent it in terms of the $CP^1$ formalism,
\begin{align} 
& S^a=\frac{1}{2}z^{\dagger}\sigma^a z,~~ z=(z_{1},~z_{2}).
\end{align}
The spatial fluctuations of the percolating domain walls trigger the hopping of dipoles, (i.e. a pair of spinons between the rungs) along the stripe transverse to their dipole moment. An important characteristic of this transition is that the percolation at the QCP does not trigger the hopping of a single spinon on each row, stemming from the fact that each domain wall contains a pair of spinons (a dipole) between the rungs. Based on these key observations, we propose that the low energy effective description of the QCP is a critical dipole liquid theory :
\begin{align} 
&\mathcal{L}=\sum_{i=1,2}\sum_{\alpha=1,2} (\partial_t \theta_i^{\alpha}+a^{\alpha}_0)^2-K (\partial_x (\theta_i^1-\theta_i^2)+a^1_{x}-a^2_x)^2,\nonumber\\
&a_{x}\rightarrow a_{x}+\partial_x \gamma, ~a_{0}\rightarrow a_{t}+\partial_t \gamma,~~
z^{\alpha}_i={n}^{\alpha}_i e^{-i \theta^{\alpha}_i},
\label{ac}
\end{align}
where the lower index $i=1,2$ labels the two components of the $CP^1$ field $(z_1,z_2),$ and the upper index $\alpha=1,2$
denotes the spinon on the first and second row of the ladder.
Here we have introduced a number-phase representation for the $CP^1$ spinons, and $a_0,a_{x}$ are components of the emergent gauge field that couple to the $CP^1$ field; their gauge transformations are also listed above. We note that we are ignoring the compactification of the boson fields $\theta_i$, and have expanded them to quadratic order~\cite{you2019emergent}. The legitimacy of this approximation, which is tied to the irrelevance of instanton tunneling events, is discussed in detail later.

To further analyze this theory, we can decompose the two $CP^1$ phase fields as $\theta^{\alpha}_{\pm}=\theta^{\alpha}_1 \pm \theta^{\alpha}_2$ to find:
\begin{align} 
&\mathcal{L}=\sum_{\alpha=1,2} (\partial_t \theta_+^{\alpha}+a^i_0)^2-K (\partial_x (\theta_+^1-\theta_+^2)+a^1_{x}-a^2_x)^2+(\partial_t \theta_-^{\alpha})^2-K (\partial_x (\theta_-^1-\theta_-^2))^2.
\label{2dspin23}
\end{align}
We find that only the $\theta^{\alpha}_+$ branch couples to the emergent gauge field $a$. Due to the intra-site XY interaction in each unit cell, any even number of unpaired spinons per unit cell can be paired into a singlet, so the spinon here is only well defined modulo 2. This is also apparent from the weak intra-site XY interaction which has a tendency to pair two spins with a coupling of the form $z^{\dagger}_1(R,1)z^{\dagger}_2(R,2)-z^{\dagger}_2(R,1)z^{\dagger}_1(R,2).$ Hence the gauge charge conservation of the spinon is broken/Higgs'ed from U(1) to $Z_2$. Subsequently, the $\theta_+$ mode in Eq.~\eqref{2dspin23} at the critical point is gapped by the Higgs-mechanism, and does not contribute to the low energy physics\footnote{Such Higgs'ing does not affect the magnon mode as the physical U(1) symmetry with respect to $S_z$ charge is not broken.}. This also implies the emergent gauge field $a_x$ carried by the spinon is Higgs'ed into a discrete $Z_2$ gauge field $a_x=0,\pi$.

The field $\theta_-$ denotes the magnon mode: $b^\dagger=e^{i \theta_-}$ that carries the $S_z$ quantum number. This field contributes to the gapless degrees of freedom at quantum criticality, and henceforth we only focus on the $\theta_-$ branch:
 \begin{align} 
&\mathcal{L}=(\partial_t \theta_-^{\alpha})^2-K (\partial_x (\theta_-^1-\theta_-^2))^2.
\label{2dspin33}
\end{align}
Due to subsystem $S_z$ conservation on each row and rung, the $\theta_-^1+\theta_-^2$ branch is gapped as the fluctuation of a single boson is prohibited by symmetry. In contrast, the $\theta_-^1-\theta_-^2$ denotes the dipole field composed of a hardcore boson pair on each rung, and  at the critical point, the dipole dynamics emerges due to the proliferation of 'domain walls' along the stripe. The resultant theory for this critical point is:
 \begin{align} 
&\mathcal{L}=(\partial_t (\theta_-^1-\theta_-^2))^2-K (\partial_x (\theta_-^1-\theta_-^2))^2,
\label{cft}
\end{align} which resembles the 1D critical boson characterized by a $c=1$ conformal field theory (CFT). The remarkable feature of this  critical theory between the HOTI and trivial Mott phases is that the critical boson is a dipole consisting of two bosons, one from each row, which fluctuate together transversely along the stripe. In comparison, the dynamics of a single boson is prohibited and gapped at low energy. Thus, the critical point
  exhibits characteristic fractonic features whose low energy dynamics are contributed by the dipole fluctuations. 

We justify our theoretical prediction of the critical theory in Eq.~\eqref{cft} by numerically calculating the entanglement entropy of the half-infinite stripe at the critical point of the model in Eq.~\eqref{hhhstripe}. For 1D critical bosons, the ground state entanglement entropy exceed the area law with a well-known logarithmic correction as $S(\xi)\sim\frac{c}{6} \ln(L)+const.$, where $L$ is the  length along the stripe, and $c$ is the central charge. For infinite systems, the stripe length has to be replaced by the correlation length $\xi$~\cite{Calabrese2004}, i.e $S(\xi)\sim\frac{c}{6} \ln(\xi)+const.$ In Fig~\ref{fig:1dphase}(c), we clearly observe the logarithmic scaling between the half-infinite stripe entanglement entropy and the correlation length.
We can evaluate the central charge of the critical point by reading off the slope in Fig.~\ref{fig:1dphase}(c) based on conformal field theory (CFT) predictions~\cite{Calabrese2004}. Indeed, we find that the central charge $c=1$.

 An alternative way to illustrate the fracton dynamics at the critical point is to microscopically analyze the magnon dynamics from the lattice model in Eq.~\eqref{hhhstripe}. When the interaction strength $\lambda$ grows, the inter-cell coupling term triggers the magnons to strongly fluctuate between unit cells. However, as the inter-cell coupling contains only a ring exchange term, it does not support single charge transport between cells, and hence the leading order dynamics is governed by the dipoles moving between cells. To corroborate this we find that the two-point correlator $\langle e^{i \theta_-^{\alpha}}(0) e^{-i \theta_-^{\alpha}}(x)\rangle$ is short-ranged at the QCP. In contrast, since we expect that the phase transition point manifests a critical dipole liquid, we calculate a four-point correlation function (i.e., a two-point function between two dipoles)~\cite{you2019emergent}:
\begin{equation}
\langle  e^{i \theta_-^{1}}(0) e^{-i \theta_-^{2}}(0) e^{-i \theta_-^{1}}(x) e^{i \theta_-^{2}}(x) \rangle 
=\frac{1}{(x)^{1/(K\pi^2)}}, 
\end{equation}
from which we see that the dipole correlation functions exhibit algebraic decay akin to the 1D critical boson.

\begin{figure}[h]
  \centering
    \includegraphics[width=0.4\textwidth]{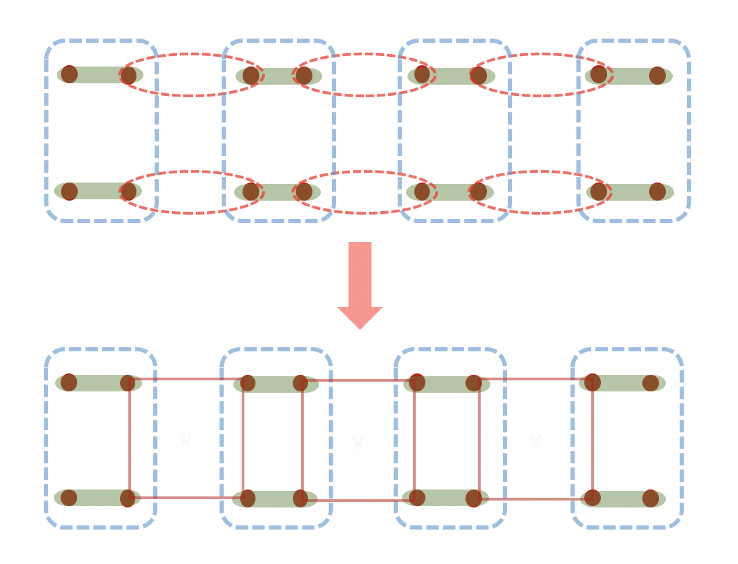} 
  \caption{Top: Two decoupled spin chains with
two spin-1/2 per unit cell (green). The XY interaction between the intra-site bond (green) and inter-site bond (red) is uniform, and the resultant state is akin to two decoupled critical chains with $c=1$ central charge.
Bottom: If we turn on strong interaction that has a tendency to fix the number of $S_z$ charge inside each blue island, the XY interaction between unit cells isre suppressed and the leading order inter-site coupling upon perturbation is the ring-exchange term on each red plaquette.} 
    \label{chain}
\end{figure}

To better understand the equivalence between this critical point and the 1D $c=1$ critical boson theory, we provide an alternative perspective from strongly coupled spin chains. Consider two identical decoupled quantum spin chains as shown in Fig.~\ref{chain}. Each chain contains a unit cell~(green bond) with
two spin-1/2's. We turn on the XY coupling between intra-, and inter-site nearest neighbor spins. When these interactions are equal, the theory forms two decoupled critical spin-1/2 chains, each with $c=1$ central charge. 
  \begin{align} 
&\mathcal{L}=\sum_{i=1,2}(\partial_t \theta_-^i)^2-K (\partial_x \theta_-^i)^2,
\label{twoboson}
\end{align}
where again the upper index $i$ refers to the chain index.
 Now we turn on inter-chain coupling on the rungs as in Fig.~\ref{chain},
\begin{align} 
&H_{int}= \sum_{i \in \text{rung}} -J (S^z_i)^2.
\end{align}
Each blue island contains four spin-1/2's from the two unit cells on each rung of the ladder. The inter-chain/rung coupling tends to fix the total $S_z$ quantum number inside each rung. This interaction plays the role of imposing a subsystem symmetry with charge conservation on each `column'.\footnote{ In the thin stripe limit, the column is just the rung connecting two chains.}
In the strong coupling limit $J \rightarrow \infty$ with fixed $S_z$ charge inside each rung, the intra-site XY coupling is not affected. However, the XY interaction between unit cells is largely suppressed as it breaks charge conservation on each rung. In the strong $J$ limit, to the leading order of perturbation theory, the inter-site coupling just contains a ring exchange term on each plaquette, and the thin stripe HOTI model from Eq.~\eqref{hhhstripe} is restored.

A particularly interesting question in this context is what happens to the effective theory of the two critical boson chains after we implement the strong inter-chain coupling $H_{int}$?
Following the duality and bosonization argument, we can map the two critical boson theories to their dual representation by expressing the boson charge density in terms of the dual vertex field $\phi^i$ (which is chosen to be an integer) as $\hat{n}^i_{-}=\partial_x \phi^i_{-}.$ As such we find:
  \begin{align} 
H_{int}=\cos((\phi^1_{-}+\phi^2_{-})2\pi).
\label{twobosondual}
\end{align}
When we fix the charge number inside each rung by adding strong $H_{int}$, the vertex field $\phi^1_{-}+\phi^2_{-}$ is pinned to be an integer value. Hence, the charge fluctuation of $\theta^1_{-}+\theta^2_{-}$ is gapped. The remaining $\theta^1_{-}-\theta^2_{-}$ branch is still gapless so the critical theory resembles a single critical boson with $c=1$ central charge. This result exactly agrees with our previous argument based on the domain wall proliferation picture.

\subsection{Berry phase with quantized dipole moment}

While the phase diagram and the critical theory between the HOTI and trivial Mott phase in the thin stripe limit can be understood using our effective theory, and complemented by the DMRG simulations, the topological signatures, e.g., a quantized many-body invariant characterizing the fingerprint of the HOTI phase is still missing. Provided that both HOTI state and the trivial Mott state share the same symmetry pattern with finite charge gaps, it is essential to search for an additional non-local observable as a smoking gun to differentiate the two phases.

In section~\ref{section1}, we proposed a quantized quadrupolarization response to characterize the HOTI phase:
\begin{align} 
&\mathcal{L}_{Q}=\frac{\Theta}{2\pi} (\partial_x \partial_y A_0-\partial_t A_{xy}).
\end{align}
In the presence of subsystem U(1) and $\mathcal{T}$ symmetry, the quadrupole moment $\Theta$ has a quantized coefficient $\Theta=0,\pi$ corresponding to the HOTI or trivial Mott phases. Such a quadrupole moment can be accessed by measuring the topological Berry phase accumulation subject to the global rank-2 flux insertion to the ground state\cite{you2019multipolar,dubinkin2019theory}.
 
For the stripe model defined in Eq.~\eqref{hhhstripe}, we can insert a global rank-2 flux $\Phi$ in the periodic cycle formed by the ladder of length $L$ by taking $A_{xy}=\Phi /L.$  This global phase insertion is achieved via the rank-2 Peierls factor $e^{i\Phi/L}$ attached to the ring-exchange terms the connect inter-site plaquettes. Now we calculate the family of ground state wavefunctions $\psi(\Phi)$ that are generated as we gradually change $\Phi$ from 0 to $2\pi$  The ground state Berry phase with respect to $\Phi$ is:
\begin{align} 
&\gamma=\int_0^{2\pi}  d\Phi  ~ \langle  \psi(\Phi)|\partial_{\Phi}|\psi(\Phi)\rangle. 
\end{align}
This Berry phase essentially measures the `quadrupole moment ' of the HOTI which concurrently manifests as the fractional corner charge at each corner, i.e., $Q_{xy}=\frac{\gamma}{2\pi}$. 

\begin{figure}[h]
  \centering
      \includegraphics[width=1\textwidth]{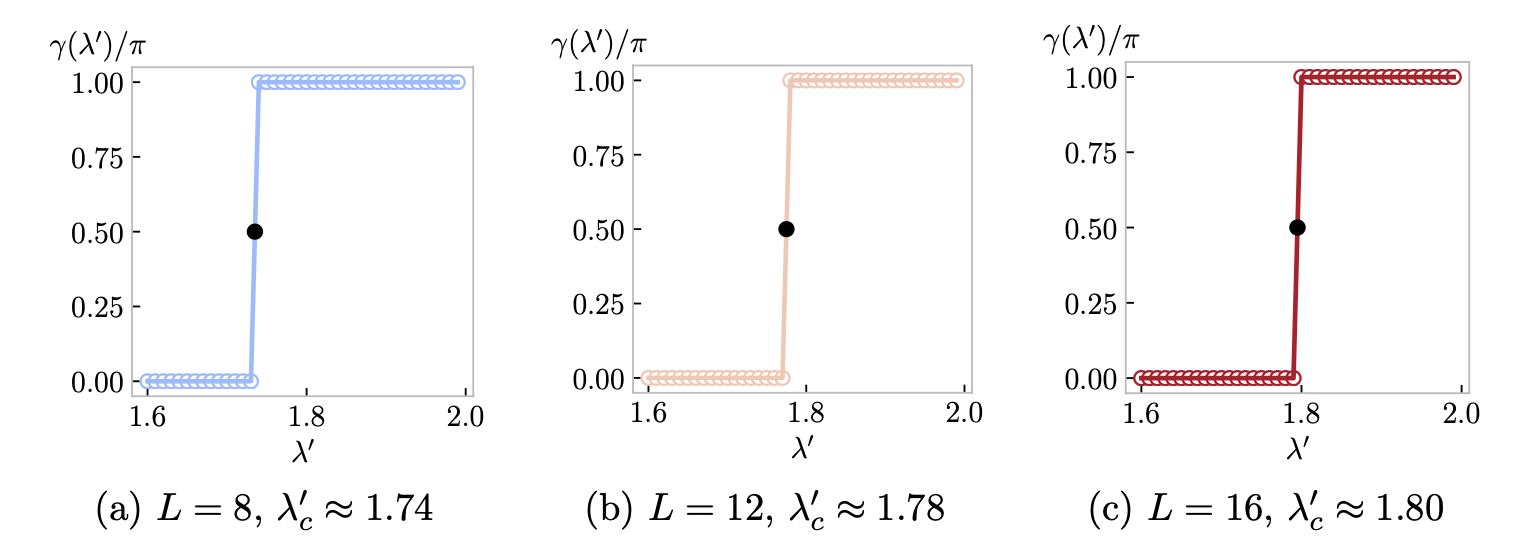} 
\caption{\textbf{Berry phase $\gamma$}. The panels show the Berry phase $\gamma(L)$ for various stripe length $L\in\{8,12,16\}$. The QCP move towards the $L\to\infty$ limit $\lambda'_c\approx 1.809$ as we increase the stripe length.}
\label{fig:berryphase}
\end{figure}

The symmetry quantization of $Q_{xy}$ guarantees that the Berry phase only takes two discrete values $\gamma=0,\pi$ as long as the ground state is gapped.
For the HOTI phase with plaquette entangled patterns, we have $Q_{xy}=1/2$ so the Berry phase accumulation is $\pi$, as is shown in Fig.~\ref{fig:berryphase}. For the trivial Mott phase, $Q_{xy}=0$ and the Berry phase accumulation is zero.
In particular, since $Q_{xy}$ is quantized to half-integers for any gapped phases with $\mathcal{T}$ and subsystem $U(1)$ symmetry, the Berry phase $\gamma$ is robust against any perturbations and hence provides a numerical signature for HOTI phases. This is exactly the feature we see in Fig. \ref{fig:berryphase} where the Berry phase remains quantized in the gapped phases, but sharply jumps by $\pi$ across the phase transition.

\section{Coupled stripe construction}
Having understood the nature of the phase transitions in the 1D ladder model, let us now get back to the phase transition theory in 2D.
Our results for the ladder indicate that the HOTI to trivial Mott phase transition in a thin stripe limit is characterized by a critical 1D boson. However, its critical behavior in 2D is still nebulous. Under normal circumstances, coupling a set of critical boson chains in a 2D wire construction can result in either a U(1) symmetry breaking instability or a transition to a Mott insulator. In particular, despite the fact that long-range order and spontaneous symmetry breaking are absent in 1D quantum systems due to the Mermin-Wagner theorem, such an instability could appear in 2D with a potential to drive the coupled 1D critical boson into a superfluid phase. Nevertheless, such a situation will not appear in our model due to the additional subsystem U(1) symmetry which allows for enhanced quantum fluctuations\cite{Nussinov2005-ab}. As the $S_z$ charge is conserved on each row and column, any single boson hopping term between cells is prevented, resulting in the absence of a symmetry breaking instability. Here and after, we will show that the 2D critical point is described by a critical dipole theory with fracton dynamics.

\begin{figure}[h]
  \centering
    \includegraphics[width=1\textwidth]{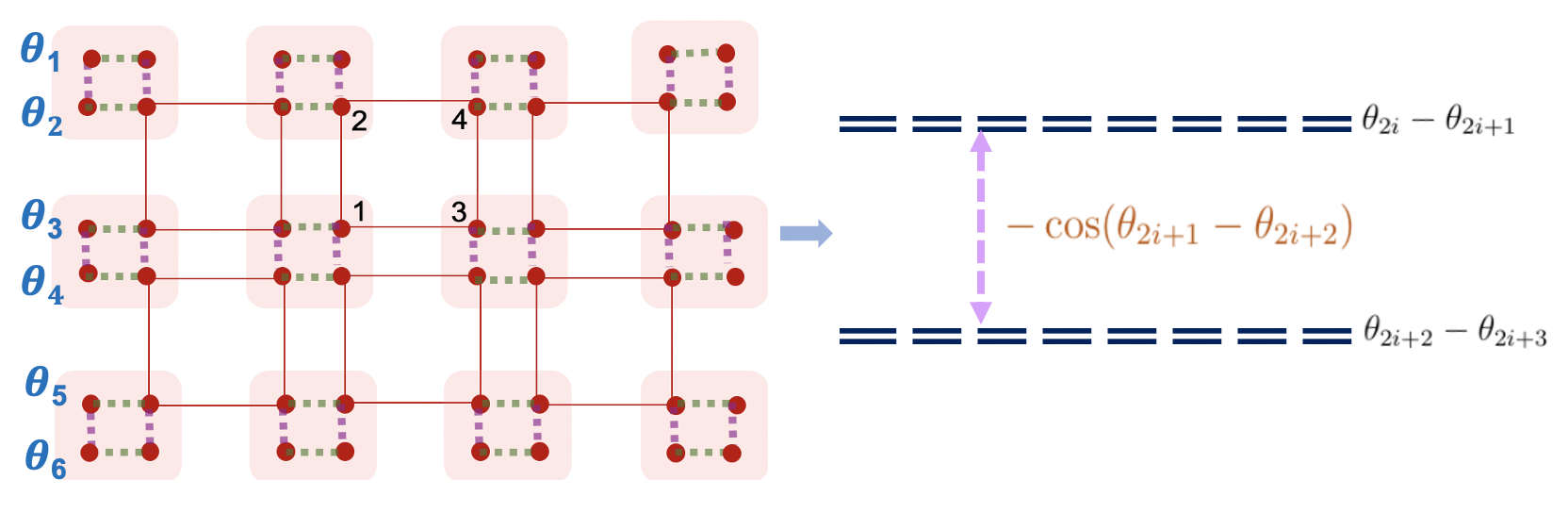} 
  \caption{Each ladder from $2i$-th and $2i+1$-th row forms a critical dipole liquid with critical dipoles oriented in $y$ that are fluctuating along the x-direction. The inter-stripe coupling triggered by the vertical XY interaction on the purple bond in each unit cell couples the stripes and tends to pin the phase of $\theta^{2i+1}_--\theta^{2i+2}_-$.} 
    \label{couple}
\end{figure}

We can extend the quasi-1D model in Eq.~\eqref{hhhstripe} to 2D by adding additional stripes along the y-direction as,
\begin{align}  
&H= \sum_{{\bf{R}}}~\left[(b^{\dagger}_{{\bf{R}},1} b_{{\bf{R}},2}+b^{\dagger}_{{\bf{R}},3} b_{{\bf{R}},4})
-\lambda'  b^{\dagger}_{{\bf{R}},2} b_{{\bf{R}}+\mathbf{e}_x,1} b^{\dagger}_{{\bf{R}}+\mathbf{e}_x+\mathbf{e}_y,4} b_{{\bf{R}}+\mathbf{e}_y,3}+h.c.\right],
\label{hhhstripe2d}
\end{align} where ${\mathbf{R}}$ runs over the sites of the 2D lattice.
Since we only kept the intra-site XY coupling along the x-bond for the 1D stripe model, the Hamiltonian (\ref{hhhstripe2d}) in 2D is merely a trivial stacking of decoupled spin ladders. 
Each ladder contains a critical dipole Luttinger liquid represented by 
\begin{align} 
\mathcal{L}=\sum^N_{i}[(\partial_t \theta^{i}_-)^2-K(\partial_x  (\theta^{2i}_--\theta^{2i+1}_-))^2].
\label{de}
\end{align}
Here $\theta_-$ denotes the phase for the hardcore boson $b,$ and the upper index labels the row number as illustrated in Fig.~\ref{couple}. Based on this definition, $e^{i(\theta^{2i}_--\theta^{2i+1}_-)}$ denotes the dipole on the rung between the $2i$-th and $(2i+1)$-th chain. Each spin ladder contains a critical dipole from the $2i$ and $2i+1$ chains fluctuating transversely along the stripe in the x-direction.

Now we turn on the  intra-site XY interaction on the y-links within each unit cell to restore the 2D model we proposed in Eq.~\eqref{hhhboson}:
\begin{align}  
&H_{\textit{coupled stripe}}= \sum_{{\bf{R}}}~(b^{\dagger}_{{\bf{R}},2} b_{{\bf{R}},3}+b^{\dagger}_{{\bf{R}},1} b_{{\bf{R}},4}).
\end{align}
Once the intra-site XY interaction on the x-links and y-links are of the same strength, the theory has $C_4$ symmetry restored, and the stripes are strongly coupled. Such a coupling, which respects all subsystem symmetries, does not change the sub-dimensional nature of the y-dipole which can only hop along the x-ladder. However, it triggers the motion of x-dipoles and lets them travel along the y-direction mediated by the inter-stripe coupling.


The stripe coupling terms in $H_{\textit{coupled stripe}}$ introduce a phase-locking term for the bosons on the y-links within each unit cell. In the bosonization language, it provide a strong coupling between the bosons from the $2i+1$-th and $2i+2$-th rows as shown in Fig.~\ref{couple}:
\begin{align} 
-\cos (\theta^{2i+1}_--\theta^{2i+2}_-).
\end{align}
Such a coupling, when sufficiently strong, will pair the two compact bosons between y-links inside each unit cell. Subsequently, we anticipate such an interaction will suppress the phase fluctuations of $\theta^{2i+1}_--\theta^{2i+2}_-$, and thus it strongly couples the stripes by imposing a constraint that $\theta^{2i+1}_-=\theta^{2i+2}_-$.
Based on this observation, one can redefine the variable $ \theta^{2i+1}_- \rightarrow \theta^i_-$ to reduce the redundant degrees of freedom, and hence the effective theory in Eq.~\eqref{de} can be written as:
\begin{align} 
\mathcal{L}=\sum^N_{i}[(\partial_t \theta^{i}_-)^2-K(\partial_x  (\theta^{i}_--\theta^{i+1}_-))^2].
\end{align}
Let us now replace the phase difference between the rows $(\theta^{i}_--\theta^{i+1}_-)$ with the
differential operator $\partial_y\theta_-(x,y,t)$ and absorb the lattice spacing into a new parameter $K$ by rescaling spacetime, and we arrive at: 
\begin{align} 
\mathcal{L}=\sum^N_{i}[K(\partial_t \theta_-)^2-K(\partial_x \partial_y\theta_-)^2].
\label{act}
\end{align} This is precisely the 2D critical dipole liquid we obtained in Ref.~\cite{you2020fracton}.
Due to subsystem $S_z$ conservation, the action in Eq.~\eqref{act} is invariant under a special U(1) transformation $
\theta_- \rightarrow \theta_- +f(x)+g(y).$
Consequently, the single magnon hopping term $(\partial_i \theta_-)^2$ is forbidden as it breaks this subsystem U(1) symmetry explicitly. Instead, the leading order dynamics originates from dipole moments oriented along the $i$-th direction, i.e., ($ \partial_i \theta_{-}$) that are constrained to move along the transverse $j$-th direction.   This is captured in our theory by the special, higher order kinetic term $(\partial_j\partial_i\theta_{-})^2$ yielding a dispersion $\omega \sim k_x k_y$.
This is remarkable because we find that the QCP exhibits characteristic fractonic features where its low energy spectrum contains both low and high-momentum modes prompted by strong locally fluctuating fields. This peculiar feature will bring about a new type of critical theory with UV-IR mixing, whose IR theory is controlled by short wave-length physics, and thus is exempt from the usual renormalization paradigm. We will come back to this point later.

   Proximate to the critical point, the sub-dimensional mobility of the boson prevents spontaneous symmetry breaking of the subsystem $U(1)$ symmetry for $S_z$ due to the Mermin-Wagner theorem~\cite{Batista2005-yr,tay2010possible,paramekanti2002ring,seiberg2020exotic}. Thus, we would not expect a superfluid instability when coupling the 2D stripes. To corroborate this we find that the two point correlator $\langle S^+(\mathbf{R}) S^-(\mathbf{R'})\rangle$ is short ranged at the QCP. Since we expect that the critical region manifests a critical dipole liquid, we find that the four-point correlation functions (two point functions between two spin-dipoles living on the same transverse stripe), exhibit an asymptotic algebraic behavior when $e_y=1, x \rightarrow \infty$\cite{you2019emergent}:
\begin{equation}
\langle S^{+}(\mathbf{R}) S^{-}(\mathbf{R}+\mathbf{e}_{y}) S^{-}(\mathbf{R}+x) S^{+}(\mathbf{R}+x+\mathbf{e}_y) \rangle
=\frac{1}{(x)^{1/(K\pi^2)}}, 
\label{dipcor}
\end{equation}
and analogously for a stripe along $y$.
These correlation functions exhibit algebraic decay and quasi-long range order if and only if the two dipoles are living on the same stripe. This again supports the fact that we should interpret the critical point as a dipole liquid having constrained subsystem dynamics. A notable property of this critical point is its dependence on spatial symmetries and lattice regularization. As we have elucidated, the quasi-long-range order only exists between the dipoles on the same stripe, and hence the critical theory breaks spatial rotation symmetry. In addition, the asymptotic behavior, including the exponent of the power-law correlator in general depends on the length of the dipole. If we further separate the distance between two charges at the end of a y-dipole, the asymptotic behavior in Eq.~\ref{dipcor} changes qualitatively and decays faster than any power-law function\cite{paramekanti2002ring,seiberg2020exotic,karch2020reduced}.

\section{Properties of the QCP}

\begin{figure}[h]
  \centering
\includegraphics[width=0.8\textwidth]{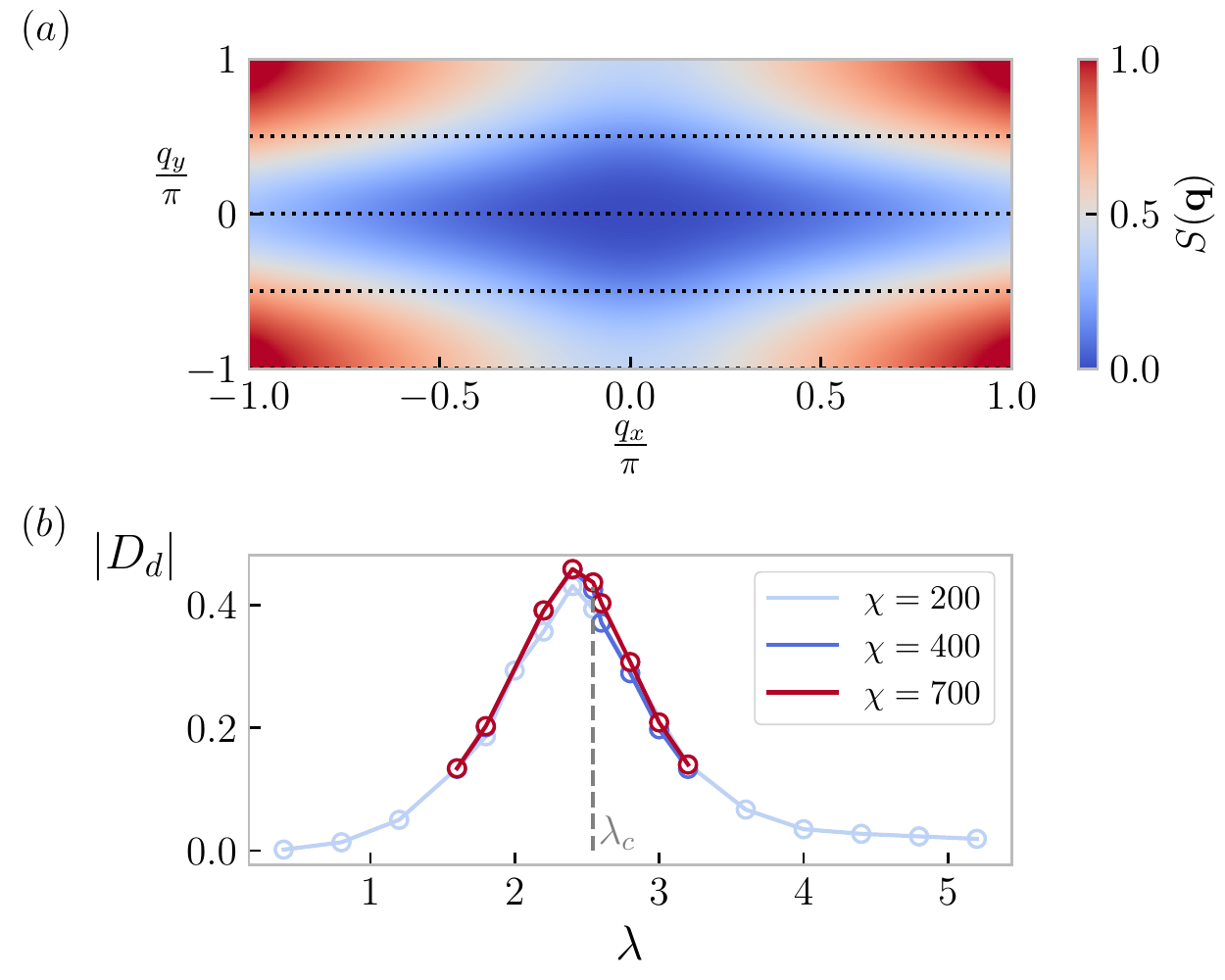}
\caption{\textbf{Static structure factor and dipole stiffness from Ref.~\cite{you2020fracton}.} $(a)$ The static structure factor $S(\mathbf{q})$ calculated for a stripe with $L_x = 100$ and $L_y = 8$ sites along the $x$ and $y$ direction, respectively. Hence, the data are obtained for momenta $q_y =0,\pm \pi,\pm \pi/2$ (black, dotted lines). $(b)$ The absolute value of the dipole stiffness $|D_d|$, calculated for an infinite cylinder with $L_y = 6$. The dipole stiffness is obtained by taking the second, symmetric derivative with step size $\delta\Phi = 0.05$.}
\label{data}
\end{figure}

The critical theory in Eq.~\eqref{act} produces a quadratic dispersion $\omega \sim k_x k_y$, which implies a `Bose surface' with characteristic lines of zero energy modes on both the $k_x$ and $k_y$ axes.  For each fixed momentum slice $k_i \neq 0$, the low energy dispersion is akin to a conventional 1D relativistic boson moving along the transverse direction. Such `quasi-1D' motion is a consequence of the sub-dimensional nature of the critical dipole liquid, i.e., the fact that  an $x(y)$-oriented dipole is only mobile along the transverse $y(x)$-oriented stripes. 
To confirm the existence of the Bose surface we numerically evaluate the structure factor:
\begin{align}
S(\mathbf{q})=&\frac{1}{N}\left[\prod^2_{m=1}\sum_{\mathbf{R}_m,\mathbf{b}_m}e^{i\mathbf{q}\cdot\left[(-1)^m\mathbf{\Lambda}_m\right]}\right]\braket{{S}^{z}_{\mathbf{\Lambda}_1}{S}^{z}_{\mathbf{\Lambda}_2}},
\end{align}
where $N=L_y L_x/4$ is the total number of unit cells, $\mathbf{\Lambda}_m = \mathbf{R}_m+\mathbf{b}_m$ specifies one the four spins $S^z_{\mathbf{R}_m,m}$ inside the unit cell $\mathbf{R}_m$ with basis vectors $\mathbf{b}_m\in\frac{1}{2}\{\mathbf{0},\mathbf{e}_x,\mathbf{e}_y,\mathbf{e}_x+\mathbf{e}_y\}$. 
In Fig.~\ref{data}(a) we show that the numerically obtained structure factor exhibits clear zero-energy lines along the $k_x,k_y$ axes. Furthermore, each $k_i \neq 0$ exhibits dispersion like a 1D relativistic boson along the transverse direction. We note that since there are nodal lines at $k_x, k_y =0,$ there is a sub-extensive number of quasi-1D modes, and the specific heat at low temperature will scale as $C_v \sim T \ln(1/T),$ which is similar to marginal Fermi liquid theory in 2D~\cite{you2019emergent,paramekanti2002ring,xu2007bond}.

As a further confirmation of the dipole liquid critical point,  we can also characterize the dipole liquid by its transport properties since our model has dipole conservation. In the language of Ref.~\cite{dubinkin2019theory}, we expect the critical dipole liquid to act as a `dipole metal' and exhibit a non-vanishing dipole conductivity in the presence of a uniform rank-2 electric field, e.g., $E_{xy}.$ In analogy to Kohn's seminal work on defining conductors and insulators~\cite{kohn1964theory}, Ref.~\cite{dubinkin2019theory} proposed a criterion to establish the existence of a dipole metal. This criterion is based on the dipole stiffness, defined as
\begin{equation}
    D_d=-\frac{\pi}{V}\frac{\partial^2 E_0}{\partial \mathfrak{q}^2},
\end{equation} where $E_0$ is the ground-state energy, and $\mathfrak{q}$ represents a constant shift of the rank-2 gauge potential $A_{xy}=\mathfrak{q}.$ The key criterion is that if $D_d$ is non-vanishing in the thermodynamic limit, then the system is a dipole metal. Our model naturally couples to $A_{xy}$ through a Peierls factor on ring-exchange terms
$
 S^{+}_{2} S^{-}_{1}S^{+}_{4} S^{-}_{3}\to e^{iA_{xy}} S^{+}_{2} S^{-}_{1} S^{+}_{4} S^{-}_{3}
$~\cite{you2019multipolar},
where we again omitted unit cell labels. Hence, we can use the sensitivity of the QCP to rank-2 boundary condition twists to determine if it is a dipole metal as we expect. Numerically, the dipole stiffness can be obtained by a two-step process: (i) we multiply each ring-exchange term by a phase factor $e^{4\pi i\frac{\Phi}{L_y}}$~\footnote{Note the number of unit cells along $y$ is equal to $L_y/2$}, (ii) we take the second, symmetric derivative of the flux-dependent ground state energy $E_0(\Phi)$ to obtain $D_d$, which is shown in Fig.~\ref{data}(b). Our numerical results clearly support our theory since the system shows a non-vanishing $D_d$ only in the neighborhood of the critical point. We note that although the parameter value of the QCP obtained via the correlation length and the dipole stiffness differ slightly from each other at $\lambda_c=2.5$, both suggest a single QCP. The deviation is a numerical artifact since DMRG at a QCP in 2D suffers not only from the effective long range couplings, but also from the enormous amount of entanglement at this particular, quantum critical point.

\section{Instability and proximate Mott phases}

Finally, we discuss the instabilities of the critical region which trigger either the proximate HOTI state or the trivial Mott phase.
The critical theory for a we proposed is given by a sum of terms of the form \begin{align} 
\mathcal{L}=\frac{K}{2}(\partial_t \theta_-)^2-\frac{K}{2}(\partial_x \partial_y \theta_- )^2
\label{theory}
\end{align} Such a quadratic Lagrangian, in which all terms involve derivatives of the fields, describes a scale invariant phase at long length scales. In a sense it can be viewed as a ``fixed point" Lagrangian. We first address the legitimacy of the spin wave approximation in arriving at Eq.~\eqref{theory}. The critical boson at the QCP is compact and the Gaussian expansion of the kinetic theory in Eq.~\eqref{act}, and hence Eq. \eqref{theory}, is valid only provided all instanton-proliferation processes that induce a $2\pi$ tunneling of the phase field $\theta_{-}$ are irrelevant.

To address this question we follow the duality and bosonization argument introduced in Ref.~\cite{xu2007bond,paramekanti2002ring}, such thatwe can map the theory to its dual representation.  As the $\theta_-$ fields are compact with the identification $\theta_-=\theta_-+2\pi Z $, particular types of topological defects will be allowed, which can be most conveniently addressed by passing to a dual representation $\hat{N},\phi$ defined on the plaquette centers\footnote{$\hat{n}_-=s_z+1/2$ is the conjugate variable of $\theta_-$.} where
\begin{align} 
\hat{n}_--1/2=\partial_i\partial_j \phi_-,~\hat{N}_-=\partial_i\partial_j \theta_-
\label{dual}
\end{align}
$\hat{N}$ and $\phi$ are a pair of conjugate variables with $\phi$ being discrete-valued, and $\hat{N}$ being compact with $N \in [0,2\pi]$. 
\begin{figure}[h]
   \centering
\includegraphics[width=0.3\textwidth]{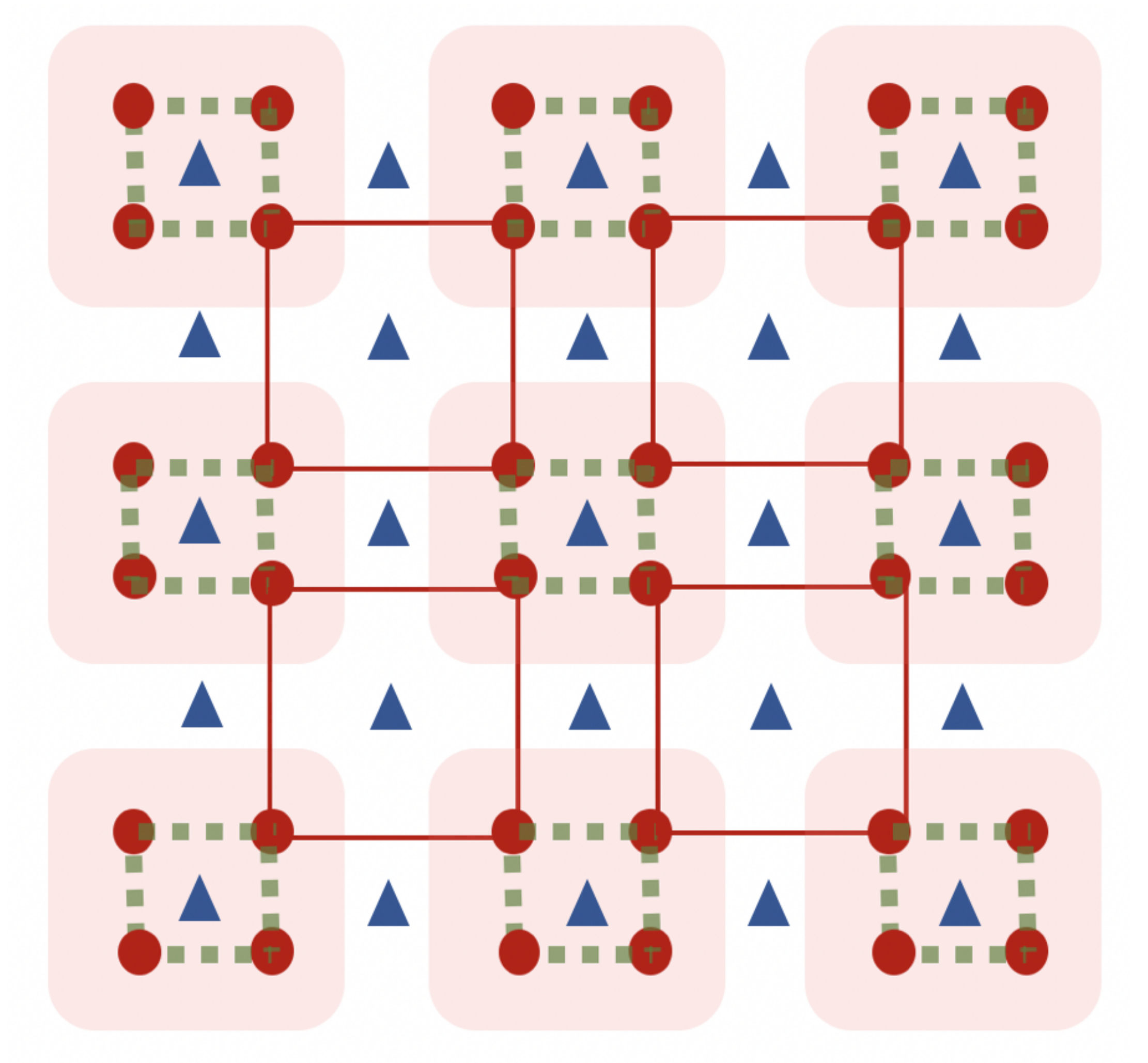}
\caption{For the dual description we enlarge each unit cell into a plaquette with 4 spins spatially separated apart so the original model can be viewed as the spin-1/2 model on a super-lattice. The blue triangles are the dual variables $\phi, \hat{N}$ living on the dual lattice.}
\label{app}
\end{figure} For simplicity, we enlarge each unit cell into a plaquette with 4 spins spatially separated apart as in Fig.~\ref{app}, so the original model can be viewed as the spin-1/2 model on a super-lattice.

The Gaussian part of the dual action is
\begin{align} 
\mathcal{L}=\frac{1}{2K}(\partial_t \phi_-)^2-\frac{1}{2K}(\partial_x \partial_y\phi_-)^2.
\label{2dspindual}
\end{align}
The $K$ can be viewed as the Luttinger liquid parameter whose value depends on the magnon interaction.
In this dual picture, due to the discreteness of $\phi_-$, one can also add vertex operators such as $\cos(4 \pi \partial_i \phi_-)$ at the QCP. We now demonstrate that as long as $K$ passes a critical value, the instanton operator is irrelevant so the spin-wave approximation is valid. Let us consider the instanton events that shift  $\theta_-$ by $2N\pi$. First, the $\mathcal{T}$ symmetry effectively sets the hardcore boson at half-filling, and in the dual representation, this implies that $\phi_-$ cannot be uniform in space. As such it is chosen to be integer and half-integer for different sites and we only consider the vertex operators $V=\cos( 4\pi \partial_i\phi_-), V'=\cos(2\pi\partial^2_i \phi_-)$ in the IR limit. In Refs.~\cite{xu2008resonating,paramekanti2002ring}, it was shown that there is a finite region for $K>K_c$ where all vertex operators are irrelevant, so the compactness of the boson can be ignored in this regime. 
As the Luttinger parameter is determined by the effective interaction, this requires the dipole to be weakly interacting, and hence the $S_z(r) S_z(r')$ interaction should be relatively small compared to the ring-exchange term and XY term. In our Hamiltonian in Eq.~\eqref{hhh}, we tune the $S_z(r) S_z(r')$ to zero so the
Luttinger parameter is large enough to escape from instanton events. It is noteworthy to emphasize that the vertex operator $(-1)^y \cos(2\pi \partial_x \phi_-),(-1)^x\cos(2\pi \partial_y \phi_-)$ with staggered sign-factor would coarse grain to zero in the long wave-length limit, and hence can be ignored in the IR theory at the QCP.

Let us also comment on the possiblity of enlarging the critical point to a critical superfluid phase. In the original representation, the correlation between two charges $\langle \cos(\theta_-(r)) \cos(\theta_-(0))\rangle$ vanishes at long-wavelength due to the subsystem $U(1)$ symmetry.  The leading order non-vanishing correlation functions are between two dipole operators living on bonds,
\begin{align} 
&\langle \cos(\partial_x\theta_-)(0,0,0) \cos(\partial_x\theta_-)(0,y,\tau) \rangle=\frac{1}{(\tau^2+y^2)^{1/(K \pi^2)}}\nonumber\\
&\langle \cos(\partial_y\theta_-)(0,0,0) \cos(\partial_y\theta_-)(x,0,\tau) \rangle=\frac{1}{(\tau^2+x^2)^{1/(K \pi^2)}}.
\end{align}
Notice that the dipole-$i$ correlation function is only nonzero when they are at the same row transverse to the dipole orientation. Thus, the dipoles effectively behave as a $1D$ Luttinger liquid with restricted motion and algebraic correlations on each stripe. This quasi-one-dimensional behavior is crucial for the existence of a critical point. As the dipole displays $1D$ motion within the same stripe, the quantum fluctuation forbids any dipole condensation with off-diagonal long-range order as a consequence of the Mermin-Wagner theorem, hence the critical point cannot be enlarged to a 2D superfluid phase.

Finally, we can consider the instabilities of the theory in Eq.~\eqref{act} which will drive it into one of the gapped phases proximate to the QCP. When tuning away from the critical point, either the inter-site plaquette-ring exchange term or the intra-site XY term can trigger a gap via instanton proliferation\cite{gogolin1999bosonization} generated by the addition of 
\begin{align}
&(\lambda-\lambda_c)[\cos(2\pi \partial_x \phi_-)+\cos(2\pi \partial_y \phi_-)],
\label{defect1}
\end{align} to the dipole liquid theory.
Here $\phi$ is the vertex operator defined as $\partial_x\partial_y \phi_-=S^z=\hat{n}_b-1/2$ which triggers the $2\pi$ instanton tunneling of the compact boson.  Such a term is induced by the imbalance between intra-site and inter-site interactions, and is absent at the fine-tuned critical point. Once away from the critical region, the instanton proliferation gaps out the critical dipole liquid and yields either an onsite entangled state as the trivial Mott phase, or an inter-site plaquette ordering as the HOTI phase.

When passing the critical point, the competition and imbalance between the onsite interaction versus the inter-site plaquette interaction triggers another vertex~(instanton proliferation) term $(\lambda-\lambda_c)[ \cos(2\pi \partial_x \phi_-)+\cos(2\pi \partial_y \phi_-)]$.
Due to $\mathcal{T}$ symmetry, which is charge-conjugation for the hardcore boson and imposes half-filling, $\phi$ cannot be chosen to be all integers on the dual sites, and at least one of them among the four sites on the dual plaquette has to be a half-integer. In addition, the translation symmetry~(here we consider the stretched lattice with each unit-cell enlarged to a plaquette as Fig.~\ref{app}) acts as,
\begin{align}
&T_x:\phi(x,y)\rightarrow \phi(x+1,y)+\frac{y}{2},~T_y:\phi(x,y)\rightarrow \phi(x,y+1)+\frac{x}{2}.
\end{align}
This implies $\cos(2\pi \partial_x \phi)+\cos(2\pi \partial_y \phi)$ breaks lattice translation (inside the conventional unit cells) and inversion and thus drives the imbalance between distinct plaquette configurations. Such an operator is absent at the QCP where the intra- and inter- unit cell coupling are balanced when  $\lambda=\lambda_c$ is fine tuned.
However, once we move away from the critical point $\lambda_c$, the UV Hamiltonian generates such vertex terms, and hence generates either a HOTI state with plaqutte order or a trivial Mott phase with on-site entangled patterns.

To visualize the relation between vertex operator and distinct plaquette patterns, we investigate the effect of vertex operator
$V_i=\cos( 2\pi \partial_i\phi_-)$. This operator creates a kink for the dipole-i along the transverse j-stripe, and hence generates a dipole gap which breaks translation $T_x,T_y,$ but is still invariant under $\mathcal{T}$. 
From either symmetry arguments, or from renormalization group flow of the lattice theory\cite{xu2007bond}, this term induces a plaquette order between two dipoles on the transverse stripe with four spins entangled at the corner of a square
\begin{align} 
&\cos( 2\pi \partial_i\phi_-)\sim\cos( 2\pi \partial_y\partial_x \theta_-)e^{i\pi r_i}.
\end{align}
Depending on the sign of $(\lambda-\lambda_c)$, the plaquette ordered pattern either falls into the unit cell or the inter-site plaquette, which exactly matches the feature of trivial Mott or HOTI phase.

It is noteworthy to emphasize that determining the relevancy and the scaling dimension of the vertex operator is considered based on the subsystem instead of the full space-time dimension. General renormalization group reasoning implies that quantum operators are irrelevant when the associated scaling dimension exceeds the space-time dimension. But due to the constrained form of the dipole correlators, which only exhibit a power-law decay on a reduced spatial region, i.e.,  on the stripe, one expects that the condition for irrelevance should be modified, and the instability should appear only provided the associated scaling dimension exceeds the subsystem space-time dimension. Subsequently, when evaluating the scaling dimension of the vertex operator $ \cos(2\pi \partial_x \phi_-)$, we only consider its scaling along the y-stripe. Renormalization on the subsystem is a significant new element in our critical theory and it implies that the IR theory allows discontinuous field configurations due to subsystem symmetry. Such discontinuous field configurations in the effective field theory are manifested by the survival of high momentum modes in the low energy due to the `Bose surface' in momentum space.

\section{Superfluid instability}

Our aforementioned discussion concentrates on the HOTI transition with subsystem U(1) symmetries. In the presence of charge conservation on subsystem rows and columns, the critical dipole liquid is stable against any charge condensation with U(1) symmetry breaking due to the subdimensional mobility of the quasiparticles. 

If we impose a strong subsystem U(1) symmetry breaking term in the Hamiltonian by introducing boson-bilinear hopping term between the sites, the QCP between HOTI and trivial phase should be enlarged into a superfluid phase which breaks charge U(1). This superfluid phase between a HOTI and a trivial Mott insulator was observed in Ref.~\cite{2019arXiv191104149B} in a boson Hubbard model with strong-weak alternating hopping term on a square lattice. When the intra/inter unit cell hopping is strong, the system stays in the trivial/HOTI phase with the four bosons mainly entangled within the enlarged unit cell or between the plaquette. When the intra/inter coupling are comparable, the theory is akin to the superfluid boson~(XY model) on a square lattice. 

Here, we highlighted an alternative situation that is intrinsically nontrivial. Suppose there is a mechanism to generate infinitesimal weak boson-bilinear hopping term between sites with weak subsystem symmetry breaking, we should ask if our critical point stable against such weak symmetry breaking instability? Explicitly, the inter-site boson hopping can be triggered by adding a tunneling term such as:
\begin{equation}
H_{\text{tunnel}}=g \cos{N \partial_i \theta_-}.
\end{equation}
 The relevancy of $g$ at the critical point determines the instability of inter-site hopping. In Ref.~\cite{xu2008resonating,paramekanti2002ring}, it was shown that there could be a finite region where the $\cos{N \partial_i \theta_-}$ operator is irrelevant, and hence where the critical point will exhibit emergent subsystem U(1) symmetry. Hence, the critical point we obtained could be immune to a certain type of weak subsystem symmetry breaking instabilities.

\section{UV-IR mixing, a critical theory beyond Renormalization}

The quantum critical points connecting different phase patterns
have broad implications for many aspects of many-body systems. The divergence of the correlation length in the critical region implies the effective interaction in the IR becomes highly non-local and hence requires us to visualize the system at a larger scaling at long wave-length. In the meantime, the divergent correlation length at the critical point can engender numerous intriguing phenomena, including emergent symmetries
that do not exist in the concrete lattice model, but emerge at low energy  near a quantum critical point. A useful description of such a  phenomenon is the renormalization group (RG) framework where emergent symmetries generally occur when the symmetry breaking terms are all irrelevant under the RG flow in the field theory description. The emergent symmetry could even engender emergent quantum anomalies at the critical point, and hence bring about new types of phase transitions beyond the GLW paradigm\cite{levinsenthil}.

Our most complete understanding of critical phenomena is accomplished 
via the RG theory. In particular, the universal properties of a wide class of interacting, random, or statistical systems can be understood by coarse-graining out the short wave-length physics due to local fluctuations, and focusing on the long wave-length behavior. Based on this observation, the critical phenomena share many universal properties that are independent of the UV Hamiltonian, but only rely the symmetry and dimensionality. For instance, the correlation functions of an operator at the critical point have a universal power-law exponent $\langle C(r) C(0) \rangle=\frac{1}{r^{D-2+\eta}}$ with $D$ being the space-time, and $\eta$ is the anomalous dimension correction. Such a universal power-law lies in the fact that the low energy part of the spectrum is contributed by long wave-length physics fluctuations at small momenta. Based on this assumption, the field patterns at low energy, which control the IR behavior, are determined merely by the spatial dimension and dynamical exponent that are insensitive to UV cut-offs.

However, the quantum critical point we demonstrated in this manuscript is a peculiar example that escapes the conventional renormalization group picture. A straightforward calculation of the density correlator at the critical point shows, for example,
\begin{align} 
&\langle S^z(r) S^z(0) \rangle=\langle n(r) n(0) \rangle=\frac{1}{r_x^2 r_y^2}.
\end{align}
The power-law correlation of the density operator at the phase transition point is spatially anisotropic with only a $C_4$ symmetry and the decay exponent does not match any known universality class. 
Motivated by this strange behavior, we can measure the asymptotic behavior of the four-boson correlator at the corner of a rectangle that respects subsystem U(1) symmetry:
\begin{align} \label{fourp}
&\langle S^+(0,0) S^-(r_x,0) S^+(r_x,r_y) S^-(0,r_y) \rangle=e^{-a (\ln(r_x) \ln(r_y))}.
\end{align}
From this we find that the double linear divergence of the energy spectrum at $k_x\rightarrow 0, k_y \rightarrow 0$ leads to the special double logarithmic scaling in real space so the
correlator shows a faster decay \cite{paramekanti2002ring,wu2020categorical} than any power-law when $r_x,r_y\rightarrow \infty$.

If we shrink the four points in Eq.~\ref{fourp} onto the corner of a thin and long stripe by 
taking $r_x=m a$ ($a$ being the lattice spacing), this four-point correlation becomes the dipole correlation function we measured in Eq.~\ref{dipcor} with a power-law decay along the y-direction. However, by changing the length of the dipole, the exponent of the correlation changes rapidly, and indicates the scaling dimension of the dipole operator depends on a UV cut-off. This phenomenon, know as UV-IR mixing, opens a new page for quantum critical theories as the low energy behavior is now sensitive to the UV degrees of freedom. In particular, the special double logarithmic scaling arises from the subsystem symmetries which produces a low energy spectrum that contains a subextensive number of exact zero-modes on the $k_x,k_y$ axes. This excessive number of low energy states at high momentum triggers a set of field configurations with strong local fluctuations at short wave-length. As opposed to the usual phase transition critical point whose low energy modes are controlled by long wave-length physics, our critical theory is dominated by short wave-length physics due to the existence of the Bose surface supporting zero modes at high momentum. These short wave-length, low energy modes engender strong local fluctuations at the critical point and hence cause the breakdown of the conventional renormalization group picture. In particular, we cannot simply integrate out (coarse grain) the local fluctuations, nor change the UV cut-off, as the high momentum modes with zero energy would bring additional singularity and hence qualitatively change the universal behavior. This is an interesting new paradigm that challenges our intuition and will be an exciting avenue for future research.

\section{Outlook}

In summary, we provided a coupled stripe construction to characterize a HOTI to trivial Mott insulator transition whose critical point displays fracton dynamics. Our result sheds new insights for various topological phase transitions and critical points beyond the GLW paradigm. Indeed, a currently popular topic in this regard is whether the critical region inherits the topological properties of the proximate higher-order topological structure, and how such a phase transition is influenced by the topological structure or entanglement pattern of the non-trivial phase. This raises important questions regarding the nature of quantum critical points between topologically distinct ground state patterns.

We also expect that the exploration of quantum critical points with subsystem symmetry can open a new page for unconventional quantum phase transitions whose low energy theory contains new classes of field configurations due to subsystem symmetry. We expect such exotic symmetry patterns can enhance quantum fluctuations and thus bring about new quantum criticality beyond our current knowledge.

{\em \textbf{Acknowledgments}~---}This work is initiated at KITP and YY, TLH, FP are supported in part by the National Science Foundation under Grant No. NSF PHY- 1748958(KITP) during the Topological Quantum Matter program.  TLH thanks the US National Science Foundation (NSF) MRSEC program under NSF Award Number DMR-1720633 (SuperSEED) and the award DMR 1351895-CAR for support.FP acknowledges the support of the DFG Research Unit FOR 1807 through grants no. PO 1370/2-1, TRR80, and the Deutsche Forschungsgemeinschaft (DFG, German Research Foundation) under Germany's Excellence Strategy EXC-2111-390814868.

\end{document}